\documentstyle[eqsecnum,floats,preprint,aps]{revtex}

\tighten

\begin{document}

\newcommand{\vp}{\varphi}
\newcommand{\CP}{${\cal CP}$}
\newcommand{\C}{${\cal C}$}
\newcommand{\Parity}{${\cal P}$}
\newcommand{\B}{${\cal B}$}
\newcommand{\be}{\begin{equation}}
\newcommand{\ee}{\end{equation}}
\newcommand{\bea}{\begin{eqnarray}}
\newcommand{\eea}{\end{eqnarray}}

\title{Electroweak Baryogenesis: A Brief Review
\footnote{Review Talk, to appear in the Proceedings of the XXXIIIrd
Rencontres de Moriond, ``Electroweak Interactions and Unified Theories'', 
March 14-21,1998, \'Editions Fronti\`eres.}}

\author{Mark Trodden\footnote{trodden@theory1.phys.cwru.edu.}}

\address{~\\Particle Astrophysics Theory Group \\
Department of Physics \\
Case Western Reserve University \\
10900 Euclid Avenue \\
Cleveland, OH 44106-7079, USA.}

\maketitle
\begin{abstract}
A brief review of the fundamental ideas behind 
electroweak baryogenesis is presented.
Since a successful implementation
of these ideas requires an extension of the minimal standard model, I 
comment on the necessary physics and how experimental constraints make these
scenarios testable at the LHC, and perhaps at existing colliders.
\end{abstract}

\setcounter{page}{0}
\thispagestyle{empty}
\vfill
\baselineskip 14pt

\noindent CWRU-P18-98

\eject

\vfill

\eject

\baselineskip 24pt plus 2pt minus 2pt

\section{Introduction}
We've heard a lot of interesting talks in this conference about electroweak 
physics at high energies in colliders. In this talk, I want to describe one
way in which electroweak physics, at high temperatures instead of energies, 
can play an important role in another arena, that of early universe cosmology.

A clear observational fact about the universe is that it is baryon-antibaryon
asymmetric. In fact, a combination of results, ranging from everyday
experience to cosmic $\gamma$-ray abundances, have demonstrated that there
is negligible primordial antimatter in our observable universe \cite{GS 76}. 
To obtain a
quantitative measure of this asymmetry we look to the standard cosmological 
model. One of the major successes of cosmology is an accurate prediction of
the abundances of all the light elements; a calculation which requires a
single input parameter, the {\it baryon to entropy ratio}

\be
\eta \equiv \frac{n_B}{s} = \frac{n_b-n_{\bar b}}{s} \ ,
\ee
where $n_b$ is the number density of baryons, $n_{\bar b}$ is that of
antibaryons, and $s$ denotes the entropy density. 
If one compares calculations
of elemental abundances with observations,
then there is agreement between these numbers if 

\be
1.5\times 10^{-10} < \eta < 7\times 10^{-10} \ .
\label{nucleo}
\ee

We {\it could} just accept this as an input parameter for the evolution of the
universe. However, it is part of the philosophy of modern cosmology to seek
an explanation for the required value of $\eta$ using quantum field theories of 
elementary particles in the early universe. While a number of different scenarios 
for generating $\eta$ have been suggested, I'm going to describe those which
make use of anomalous physics at the electroweak scale. These scenarios are 
collectively referred to as {\it electroweak baryogenesis}, and have the virtue
of relying on physics that is testable at terrestrial colliders. Because
of space constraints I'll only be able to describe the basic picture, but I
hope to be able to give a feel for the fundamental physics and challenges
involved. Also, it is impossible to 
correctly reference a short article on a huge subject such as this. I have
therefore only used references where I feel they are crucial, and hope that
my colleagues will understand. For more detailed accounts of the subject and
more complete referencing of the material, I 
refer the reader to existing longer reviews \cite{reviews}.

\section{The Sakharov Criteria}
If we're going to use a particle physics model to generate the baryon asymmetry
of the universe (BAU), what properties must the theory possess? This question
was first addressed by Sakharov \cite{Sakharov} in 1967, resulting in the
following criteria

\begin{itemize}
\item Violation of the baryon number ($B$) symmetry.

\item Violation of the discrete symmetries $C$ (charge conjugation)
      and $CP$ (the composition of parity and $C$)

\item A departure from thermal equilibrium.

\end{itemize}
Of course, the first of these is obvious - no $B$ violation, no baryon production.
To understand the second condition, note that, roughly speaking, 
if $C$ and $CP$ are conserved, the
rate for any process which generates baryons is equal to that for the conjugate 
process, which produces antibaryons, so no net excess is generated on average.
Finally, in thermal equilibrium the number density of a particle species is
determined purely by its energy, and since the masses of particle and antiparticle
are equal by the CPT theorem, the number density of baryons equals that of
antibaryons.

While there exist scenarios of baryogenesis which employ putative physics at
very high energies, the central point of {\it electroweak} baryogenesis
is that all three Sakharov conditions are satisfied within the relatively
well-understood Glashow-Salam-Weinberg theory, as I'll explain.

\subsection{Baryon Number Violation}
In the standard electroweak theory baryon number is an exact global symmetry.
However, as realized by 't Hooft \cite{tHooft76}, baryon number is violated
at the quantum level through nonperturbative processes. These effects are
closely related to the nontrivial vacuum structure of the electroweak theory.
To see this, note two facts about the electroweak theory. First, one may write
a vectorlike current for baryons as

\be 
j_B^{\mu} = \frac{1}{2} {\bar Q}\gamma^{\mu}Q \ ,
\label{baryoncurrent}
\ee
where $Q$ represents quarks, and there is an implied sum over the color and 
flavor indices. Now, due to quantum effects, any axial current 
${\bar \psi}\gamma^{\mu}\gamma^5 \psi$ of a gauge 
coupled Dirac fermion $\psi$, is anomalous \cite{AdlerBJ 69}. 
This is relevant to baryon number since the electroweak fermions couple 
chirally to the gauge fields. If we write the baryon current as

\be
j_B^{\mu} = \frac{1}{4}\left[{\bar Q}\gamma^{\mu}(1-\gamma^5)Q
+{\bar Q}\gamma^{\mu}(1+\gamma^5)Q\right] \ ,
\ee
only the axial part of this vector current is important when one calculates 
the divergence. This effect can be 
seen by calculating triangle graphs and leads
to the following expressions for the divergences of the baryon number and
lepton number currents;

\be
\partial_{\mu}j_B^{\mu} = \partial_{\mu}j_l^{\mu}
=n_f\left(\frac{g^2}{32\pi^2}W_{\mu\nu}^a {\tilde W}^{a\mu\nu}
-\frac{g'^2}{32\pi^2}F_{\mu\nu}{\tilde F}^{\mu\nu}\right) \ ,
\label{anomaly}
\ee
where $W_{\mu\nu}$ is the SU(2) field strength tensor and, for simplicity,
I've ignored the U(1) interactions. Also, $n_f$ is the number of families, 
and a tilde denotes the dual tensor.

Now, the vacua of the theory may be labelled by the
{\it Chern-Simons number}, defined by

\begin{equation}
N_{CS}(t) \equiv \frac{g^2}{32 \pi^2}\int d^3x\, \epsilon^{ijk}
{\rm Tr}\left( A_i \partial_j A_k + \frac{2}{3}ig A_i A_j A_k \right)\ .
\label{NCSdef}
\end{equation}
Although this number is not gauge-invariant, we shall only be interested in 
changes in it, which are gauge-invariant. 

For simplicity, consider space to be a 3-sphere and consider the change in
baryon number from time $t=0$ to some arbitrary final time $t=t_f$.
For transitions between vacua, the change in baryon number may be written
using (\ref{NCSdef}) as

\be
\Delta B = \Delta N_{CS} \equiv n_f[N_{CS}(t_f) - N_{CS}(0)]\ .
\ee
So, the change in baryon number is associated with a change in the vacuum state 
of the system.

At zero temperature, baryon number violating events are exponentially suppressed.
This is because there exists a potential barrier between vacua and anomalous
processes are thus tunneling events. The relevant barrier height is set by the
point of least energy on the barrier. This point is known as the 
{\it sphaleron}, and has energy $E_{sph}\sim 10$ TeV.
However, at temperatures above or comparable to the critical temperature
of the electroweak phase transition, vacuum transitions over the sphaleron 
may occur frequently due to thermal activation.

Detailed calculations of the baryon number violating rate in the broken 
\cite{CM 90} and unbroken \cite{ASYi}
phases around the critical temperature, taking into account fluctuations
around the sphaleron and other effects of nonzero temperature have been performed, yielding

\begin{equation}
\Gamma(T)  = \left\{ \begin{array}{ll}
         \mu\left(\frac{M_W}{\alpha_W T}\right)^3M_W^4 
\exp\left(-\frac{E_{sph}(T)}{T}\right)   & \ \ \ \mbox{$0\ll T < T_c$} \\
         \kappa\alpha_W(\alpha_W T)^4 & \ \ \ \mbox{$T > T_c$}
         \end{array} \right. \ ,
\label{rate}
\end{equation}
where $\mu$ is a dimensionless constant. Here, the temperature-dependent
``sphaleron'' energy $E_{sph}(T)$ is defined through the finite temperature 
effective potential. The important point here is that $\Gamma(T)$ is large
for $T>T_c$ and extremely small for $T<T_c$.

\section{C and CP Violation}
Fermions in the electroweak theory are chirally coupled to the gauge fields. 
In terms of the discrete symmetries of the theory,
these chiral couplings result in the electroweak theory being maximally
C-violating.
However, the issue of CP-violation is more complex.

CP is known not to be an exact symmetry
of the weak interactions, and is observed experimentally in the neutral 
Kaon system through $K_0$, ${\bar K}_0$ mixing. Although at
present there is no completely satisfactory theoretical explanation 
of this, CP violation is a natural feature of the
standard electroweak model. The Kobayashi-Maskawa (KM) quark mass mixing
matrix contains a single
independent phase, a nonzero value for which signals CP violation.
While this is encouraging for baryogenesis, it turns out that this particular source of
CP violation is not strong enough. The relevant effects are parametrized by
a dimensionless constant which is no larger than $10^{-20}$. This appears
to be much too small to account for the observed BAU and so it is usual to turn
to extensions of the minimal theory.

There are two principal ways of doing this

\begin{itemize}
\item The Two-Higgs Doublet Model. Here there are two scalars
$\Phi_1$ and $\Phi_2$, and the scalar potential is 
replaced by the most general renormalizable two-Higgs potential.
To make the CP-violation explicit, we write the Higgs fields in
unitary gauge as

\begin{equation}
\Phi_1=(0, \varphi_1)^T \ \ \ \ , \ \ \ \ \Phi_2=(0, \varphi_2 e^{i\theta})^T
\end{equation}
where $\varphi_1$, $\varphi_2$, $\theta$ are real, and $\theta$ is the CP-odd phase.

Changes in $\theta$ are
dependent on changes in the magnitude of the Higgs fields. In particular, if
a point in space makes a transition from false electroweak vacuum to true
then $\Delta \theta >0$, and sphaleron processes result in the preferential
production of baryons over
antibaryons. For the opposite situation $\Delta \theta <0$, and sphaleron
processes generate an excess of antibaryons. 
The total change in the phase $\theta$ 
(from before the phase transition to $T=0$) is the quantity that enters into 
estimates of the BAU.

\item Higher Dimension Operators. If we view the
model as an effective field theory, valid at energies below some mass
scale $M$, one expects extra, nonrenormalizable operators, some of which will 
be CP odd. A particular dimension six example is 
\cite{misha 88}

\be
{\cal O}=\frac{b}{M^2}\,\hbox{Tr}(\Phi^{\dagger}\Phi)
\hbox{Tr}(F_{\mu\nu} \tilde{F}^{\mu\nu}) \ ,
\label{newop}
\ee
with $b$ a dimensionless constant.
\end{itemize}

Whatever its origin, the effect of CP violation on anomalous baryon number 
violating processes is to provide a fixed direction for the net change in 
baryon number.

\section{The Electroweak Phase Transition}
The question of the order of the electroweak phase transition is central to
electroweak baryogenesis. Since the equilibrium description of particle 
phenomena is extremely accurate at electroweak temperatures, baryogenesis 
cannot occur at such low scales without the aid of phase transitions.

For a continuous transition, the associated departure from
equilibrium is insufficient to lead to relevant baryon number production
\cite{KRS}. The order parameter for the electroweak phase transition is
$\varphi$, the modulus of the Higgs field.
For a first order transition the extremum at $\varphi=0$ becomes separated
from a second local minimum by an energy barrier.
At the critical temperature $T=T_c$ both phases are equally 
favored energetically and at later times the minimum at $\varphi \neq 0$ becomes
the global minimum of the theory. Around $T_c$ quantum tunneling
occurs and nucleation of bubbles of the true vacuum 
in the sea of false begins. At a particular temperature below $T_c$, bubbles
just large enough to grow nucleate. These are termed {\it critical} bubbles,
and they expand, eventually filling all of space and completing the transition.
As the bubble walls pass each point in space, the order
parameter changes rapidly, as do the other fields and this leads to a
significant departure from thermal equilibrium. Thus, if the phase 
transition is strongly enough first order it is possible to satisfy
the third Sakharov criterion in this way.

There is a further criterion to be satisfied. As the wall passes a
point in space, the Higgs fields evolve rapidly and the Higgs VEV changes from
$\langle\phi\rangle=0$ in the unbroken phase to

\be
\langle\phi\rangle=v(T_c)
\label{vatTc}
\ee
in the broken phase. Here, $v(T)$ is the value
of the order parameter at the symmetry breaking global minimum of the finite 
temperature effective potential. 
Now, CP violation and the departure from equilibrium occur while the Higgs field 
is changing. Afterwards, the point is
in the true vacuum, baryogenesis has ended, and baryon number violation
is exponentially supressed. Since baryogenesis is now over, 
it is
imperative that baryon number violation be negligible at this temperature in
the broken phase, otherwise any baryonic excess generated will be
equilibrated to zero. Such an effect is known as {\it washout} of the 
asymmetry and the criterion for this not to happen may be written as

\be
\frac{v(T_c)}{T_c} \geq 1 \ .
\label{washout}
\ee
Although there are a number of nontrivial steps
that lead to this simple criterion, (\ref{washout}) is traditionally used 
to ensure that the baryon asymmetry survives after
the wall has passed.
It is necessary that this criterion be satisfied for any electroweak 
baryogenesis scenario to be successful.

In the minimal standard model, in which the Higgs mass is now constrained 
\cite{deJong} to be $m_H > 89.3$ GeV, it is clear from numerical simulations
\cite{KLRS3 96} that (\ref{washout}) is not satisfied. This is therefore a 
second
reason to turn to extensions of the minimal model.

\section{Mechanisms}
Historically, the ways in which baryons may be produced as a bubble wall, or
phase boundary, sweeps through space, have been separated into two
categories. 

\begin{enumerate}
\item {\it local baryogenesis}: baryons are produced when the baryon number 
violating processes and CP violating processes occur together near 
the bubble walls. 
\item {\it nonlocal baryogenesis}: particles undergo CP violating
interactions with the bubble wall and carry an asymmetry in a quantum number 
other than baryon number into the unbroken phase region away from the wall.
Baryons are then produced as baryon number violating processes convert the
existing asymmetry into one in baryon number.
\end{enumerate}
In general, both local and nonlocal 
baryogenesis will occur, and the BAU will be the sum of that generated by the 
two processes.
I'll only have time to discuss one mechanism here and I'll choose 
nonlocal baryogenesis and refer anyone interested to a recent discussion of
local baryogenesis \cite{LRT 97}.

Nonlocal baryogenesis typically involves the interaction of the 
bubble wall with the various fermionic species in the unbroken phase.
The main picture is that as a result of CP violation in the bubble wall, 
particles with opposite chirality interact differently with the wall,
resulting in a net injected chiral flux. This flux
thermalizes and diffuses into the unbroken phase where it is
converted to baryons. In what follows I'll just outline a simple 
nonlocal calculation \cite{JPT2 94}.

First assume that the Higgs fields change in a narrow region at the face of
the bubble wall. We call this 
the {\it thin wall regime}, and it is valid if the mean free path 
$l$ of the fermions being considered is much greater
than the thickness $\delta$ of the wall. I'll also ignore interactions in 
the broken phase, assuming here that baryon number violation turns off 
instantly after the wall passes.

The equation for the generation of baryon number may be written as

\begin{equation}
\frac{d n_B}{dt} = -\frac{n_f\Gamma(T)}{2T}\sum_i\mu_i \ ,
\label{nonlocal B}
\end{equation}
where the rate per unit volume for electroweak sphaleron transitions 
is given by~(\ref{rate}) for $T>T_c$.
Here, $n_f$ is again the 
number of families and $\mu_i$ is the chemical potential for left 
handed particles of species $i$. The crucial question in applying this 
equation is an accurate evaluation of the chemical potentials that bias 
baryon number production. To be concrete, I shall focus on leptons 
\cite{JPT2 94}. If there is local thermal
equilibrium in front of the bubble walls - as I am assuming - then the
chemical potentials $\mu_i$ of particle species $i$ are related to
their number densities $n_i$ by
   
\begin{equation}  
n_i = {{T^2} \over {12}} k_i \mu_i \ ,
\end{equation}
where $k_i$ is a statistical factor which equals $1$ for fermions and $2$ 
for bosons.

The source term 
in the diffusion equation is the flux $J_0$ resulting from the 
asymmetric reflection and transmission of left and right handed leptons 
off the bubble wall. For left-handed leptons, the relevant flux is

\begin{equation}
J_0 \simeq {{v m_l^2 m_H \Delta \theta_{CP}} 
\over {4 \pi^2}} \ .
\end{equation}
where $v$ is the speed of the wall, $m_l$ is the lepton mass and 
I've written $\Delta\theta_{CP}$ to parameterize the CP-violation.
Now, we feed $J_0$ into the diffusion equation for a single particle species

\begin{equation}
D_L L_L^{\prime \prime} + v L_L^{\prime} = 
\xi_L J_0 \delta (z) \ ,
\label{diffusion}
\end{equation}
where $D_L$ is the diffusion constant for leptons, $\xi^L$ 
is the {\it persistence length} of the current, and a
prime denotes the spatial derivative in the direction $z$   
perpendicular to the wall. This equation may be solved to give

\begin{equation}
L_L(z) = \left\{ \begin{array}{ll}
          J_0 {{\xi_L} \over {D_L}} e^{- \lambda_D z}  & \ \ \ \ z>0 \\
          0 & \ \ \ \ z<0
         \end{array} \right. \ ,
\end{equation}
with the diffusion root 

\be
\lambda_D = \frac{v}{D_L} \ . 
\ee
In the massless approximation the chemical potential $\mu_L$ can be
related \cite{JPT2 94} to $L_L$ by 

\begin{equation}
\mu_L = { 6 \over {T^2}} L_L
\end{equation}
Inserting the sphaleron rate and the above results for the chemical potential 
$\mu$ into~(\ref{nonlocal B}), the final baryon to entropy ratio becomes

\begin{equation}
\frac{n_b}{s} = \frac{1}{4 \pi^2} \kappa 
\alpha_W^4 (g^*)^{-1}
\Delta \theta_{CP} \left(\frac{m_l}{T}\right)^2 
\frac{m_H}{\lambda_D} \frac{\xi^L}{D_L} \ ,
\end{equation}

In some models this can be large enough to explain the BAU. However, the 
calculation must be carried through in detail separately for each model.

\section{Extensions of the MSM}
As I mentioned earlier, the relevant extensions can be described roughly
as

\begin{itemize}
\item Extra light scalars coupled to the Higgs particle, that lead to a
more strongly first order phase transition and hence a strong departure
from equilibrium and negligible washout.
\item Two Higgs doublet models to get more CP-violation and enhance the phase
transition.
\item Higher dimension operators to provide extra CP-violation.
\end{itemize}

It is important to note that in electroweak baryogenesis, we may use particle
physics to constrain these extensions. For example, 
the operator ${\cal O}$ induces electric dipole moments
for the electron and the neutron, and the strongest 
experimental constraint on the size of such an operator comes from the
fact that such dipole moments have not been observed.
Working to lowest order (one-loop) \cite{LRT 97}

\begin{equation}
\frac{d_e}{e} = \frac{m_e \sin^2(\theta_W)}{8\pi^2}\frac{b}{M^2}
\ln\left(\frac{M^2 + m_H^2}{m_H^2}\right)\ .
\label{dipole}
\end{equation}
Then, using the experimental limit \cite{commins} yields the bound

\begin{equation}
\frac{b}{M^2}\ln\left(\frac{M^2 + m_H^2}{m_H^2}\right) <
\frac{1}{(3 {\rm ~TeV})^2}\ .
\label{bound}
\end{equation}
Therefore, any baryogenesis scenario which relies on CP violation introduced
via the operator ${\cal O}$ must respect the bound (\ref{bound}).

\section{The MSSM}
Here I just want to comment on how the minimal supersymmetric standard
model is at present a viable and testable candidate theory for electroweak
baryogenesis. 

In the MSSM there are two Higgs fields. At one loop, a CP-violating
interaction between these fields is induced through supersymmetry
breaking. Alternatively, there also exists extra CP-violation through
CKM-like effects in the chargino mixing matrix. Thus, there seems to be
sufficient CP violation for baryogenesis to succeed.

Now, the two Higgs fields combine to give one light scalar Higgs $h$ such
that $m_h < 125$GeV. In addition, there are also light {\it stops} (the
superpartners of the top quark) in the theory. These light scalar
particles can lead to a strongly first order phase transition if the 
scalars have masses in the correct region of parameter space. A detailed
two loop calculation \cite{CQW 97}, and lattice results
\cite{LR 98} indicate that the allowed region is given by

\bea
75 {\rm GeV}\leq & m_h & \leq 105 {\rm GeV} \\
100 {\rm GeV}\leq & m_{\tilde t} & \leq m_t \ ,
\label{MSSMconstraints}
\eea  
for $\tan\beta \equiv \langle \Phi_2 \rangle/\langle \Phi_1 \rangle \sim 2$.
In the next few years, experiments at LEP should probe this range of Higgs 
masses and we should know if the MSSM is a good candidate for electroweak
baryogenesis.

\section{Summary}
I hope I've described how electroweak baryogenesis combines beautiful
nonperturbative field theory with cosmology to possibly explain the BAU.
By necessity, the account I've given is simplified and in reality a great
deal of work has gone into understanding the relevant physics.

There exist several viable scenarios at present and, interestingly, there 
exists a window of parameter space in the MSSM that remains open, but that will
be probed soon.

Ultimately, much of the physics of electroweak baryogenesis is testable in 
colliders, and it is possible that, with the next generation of
colliders, we will understand where the matter we
are made of came from.

\vspace{1.5cm}
\begin{flushleft}
{\bf Acknowledgements}
\end{flushleft}
\vspace{5mm}

I would like to thank the organizing committee for inviting me to speak, and
for working so hard to make the conference run smoothly. 
This work was supported by the U.S. department of Energy, the National
Science Foundation (NSF) and by funds provided by Case Western Reserve
University. I am also grateful for a grant from the European Union 
Training and Mobility of Researchers Programme, and for additional help
from the NSF.

\end{document}